\documentclass[aip,jcp,floatfix,citeautoscript,preprint,11pt]{revtex4-1}
\setcitestyle{super}

\usepackage{verbatim}

\usepackage{graphicx}
\usepackage{subfigure}
\usepackage{setspace}[1.5]
\usepackage{amsmath, amsthm, amssymb}
\usepackage{xfrac}
\usepackage{dcolumn}
\usepackage{gensymb}
\usepackage{threeparttable}

\usepackage{color}
\usepackage[normalem]{ulem}

\newcommand{\etal}{\textit{et al.}}

\newcommand{\xc}{\textit{xc}~}

\newcommand{\csix}{$\text{C}_6^{\text{AB}}$}
\newcommand{\ciso}{$\text{C}_6^{\text{iso}}$}
\newcommand{\cfull}{$\text{C}_6^{\text{full}}$}

\newcommand{\alphiso}{$\alpha^{0}_{\parallel}$}
\newcommand{\alphfull}{$\alpha^{0}_{\perp}$}

\begin{document}

\date{\today} \title{Properties of the water to boron nitride
  interaction: from zero to two dimensions with benchmark accuracy}


\author{Yasmine S. Al-Hamdani} \affiliation{Thomas Young Centre and
  London Centre for Nanotechnology, 17--19 Gordon Street, London, WC1H
  0AH, U.K.}  \affiliation{Physics and Materials Science Research
  Unit, University of Luxembourg, L-1511 Luxembourg}

\author{Mariana Rossi} \affiliation{Fritz Haber Institute of the Max
  Planck Society, Faradayweg 4-6, 14195 Berlin, Germany}

\author{Dario Alf\`{e}} \affiliation{Thomas Young Centre and London
  Centre for Nanotechnology, 17--19 Gordon Street, London, WC1H 0AH,
  U.K.}  \affiliation{Department of Earth Sciences, University College
  London, Gower Street, London WC1E 6BT, U.K.}

\author{Theodoros~Tsatsoulis} \affiliation{Max Planck Institute for Solid State
Research, Heisenbergstra{\ss}e 1, D-70569 Stuttgart, Germany}

\author{Benjamin Ramberger} \affiliation{University of Vienna, Faculty
  of Physics and Center for Computational Materials Sciences,
  Sensengasse 8/12, 1090 Wien.}

\author{J. Gerit Brandenburg} \affiliation{Thomas Young Centre and London Centre
for Nanotechnology, 17--19 Gordon Street, London, WC1H 0AH, U.K.}
\affiliation{Department of Physics and Astronomy, University College London, 20
Gordon Street, London, WC1H 0AJ, U.K.} \affiliation{Department of Chemistry,
University College London, 20 Gordon Street, London WC1H 0AH, United Kingdom}

\author{Andrea Zen} \affiliation{Thomas Young Centre and London Centre
  for Nanotechnology, 17--19 Gordon Street, London, WC1H 0AH, U.K.}
\affiliation{Department of Physics and Astronomy, University College
  London, 20 Gordon Street, London, WC1H 0AJ, U.K.}

\author{Georg Kresse} \affiliation{University of Vienna, Faculty of Physics and
Center for Computational Materials Sciences, Sensengasse 8/12, 1090 Wien.}

\author{Andreas Gr\"{u}neis}\affiliation{Max Planck Institute for Solid State
Research, Heisenbergstra{\ss}e 1, D-70569 Stuttgart, Germany}

\author{Alexandre Tkatchenko} \affiliation{Physics and Materials
  Science Research Unit, University of Luxembourg, L-1511 Luxembourg}

\author{Angelos Michaelides} \affiliation{Thomas Young Centre and
  London Centre for Nanotechnology, 17--19 Gordon Street, London, WC1H
  0AH, U.K.}  \affiliation{Department of Physics and Astronomy,
  University College London, 20 Gordon Street, London, WC1H 0AJ, U.K.}

\begin{abstract} Molecular adsorption on surfaces plays an important part in
catalysis, corrosion, desalination, and various other processes that are
relevant to industry and in nature. As a complement to experiments, accurate
adsorption energies can be obtained using various sophisticated electronic
structure methods that can now be applied to periodic systems. The adsorption
energy of water on boron nitride substrates, going from zero to 2-dimensional
periodicity, is particularly interesting as it calls for an accurate treatment
of polarizable electrostatics and dispersion interactions, as well as posing a
practical challenge to experiments and electronic structure methods. Here, we
present reference adsorption energies, static polarizabilities, and dynamic
polarizabilities, for water on BN substrates of varying size and dimension.
Adsorption energies are computed with coupled cluster theory, fixed-node quantum
Monte Carlo (FNQMC), the random phase approximation (RPA), and second order
M{\o}ller-Plesset (MP2) theory. These explicitly correlated methods are found to
agree in molecular as well as periodic systems. The best estimate of the
water/h-BN adsorption energy is $-107\pm7$ meV from FNQMC. In addition, the
water adsorption energy on the BN substrates could be expected to grow
monotonically with the size of the substrate due to increased dispersion
interactions but interestingly, this is not the case here. This peculiar finding
is explained using the static polarizabilities and molecular dispersion
coefficients of the systems, as computed from time-dependent density functional
theory (DFT). Dynamic as well as static polarizabilities are found to be highly
anisotropic in these systems. In addition, the many-body dispersion method in
DFT emerges as a particularly useful estimation of finite size effects for other
expensive, many-body wavefunction based methods. \end{abstract}

\maketitle

\section{Introduction} Molecular adsorption on surfaces is a fundamentally
important process in catalysis, gas storage, water purification and many other
areas. Water especially, is ubiquitous and in the absence of an ultra high
vacuum (UHV), materials inevitably come into contact with it, which can have a
substantial impact on surfaces both in industry and in nature. Even the gecko
which is known for defeating gravity and climbing walls by adhesion from weak
van der Waals (vdW) interactions\cite{Autumn}, has been found to lose its grip
when its toepads become wet\cite{Stark3080,stark2013surface,badge2014role}.
Adsorption on so-called vdW materials is particularly exciting as they exhibit
peculiar long-range correlation interactions compared to bulk solids or
molecules\cite{gobre2013scaling}. In addition, recent efforts to treat unclean
water have involved using low dimensional materials such as graphene and
hexagonal boron nitride (h-BN) to adsorb harmful
impurities\cite{lei_porous_2013,yusuf2015applications}. vdW interactions also
play an integral role in forming complex molecular crystal structures of
compounds used in medicine\cite{price2014predicting}. In order to understand and
design new systems in this broad range of applications, an accurate knowledge of
adsorption energies is useful and often essential.

Accurate adsorption energies on clean surfaces are difficult to obtain
experimentally. That is partly because of the pristine surfaces under UHV that
have to be prepared, and secondly because few adsorption measurement techniques
exist with the level of precision that is required to measure weak physisorption
energies. Amongst those, single crystal adsorption calorimetry (SCAC) is a
particularly elegant technique\cite{brown1998femtomole,campbell2013enthalpies}
but even with this, no adsorption energy for water on 2-dimensional (2D)
surfaces has been reported yet. Fortunately, a few computational methods have
emerged over the years that can be applied to periodic systems, as useful and
reliable ways to calculate molecular adsorption energies on extended surfaces.
These include the random phase approximation (RPA), diffusion Monte Carlo (DMC)
and quantum chemical methods such as second order M{\o}ller-Plesset (MP2) theory
and coupled cluster with single, double and perturbative triple excitations
(CCSD(T)).  These electronic structure methods have had numerous successes
including for example, the RPA prediction of the adsorption site for CO on metal
surfaces\cite{schimka2010accurate} where generalised gradient approximations
(GGA) within density functional theory (DFT) fail. Another example is the
agreement between DMC and embedded coupled cluster theory for the surface
energy\cite{Binnie2010} and water adsorption energy\cite{Tsatsoulis2017} on the
LiH crystal - a material that spontaneously oxidises in moist air making any
experimental measurement extremely difficult. Theoretically calculated reference
adsorption energies are therefore helpful to experimentalists as well as to
developers of computational methods.

A key challenge is having accurate electronic structure theories that we can
solve for realistic extended surface models. However, thanks to the great
improvements in codes and algorithms, surface adsorption problems have become
increasingly accessible in recent
years\cite{Ma_11a,Voloshina_11,Beate2009,Muller2008,Muller2009,al-hamdani2,ganesh_binding_2014,gordillo_h_2013,Lei2016,Ma_11b,schimka2010accurate,binnie2011ab,wu2015interaction,wu2016hexagonal,jenness2010benchmark,KaoliniteQMC}.
Water adsorption on graphene is an exemplary case that has been computed with
the RPA, DMC\cite{Ma_11a} and embedded
CCSD(T)\cite{Voloshina_11,jenness2010benchmark}. However the reported adsorption
energies range by up to 40$\%$; likely because the calculations involved a
number of limitations and approximations which have not been quantified. Gauging
the impact from different approximations is evidently desirable and would
deliver greater understanding of the applicability of these methods.

In this study, we focus on the interaction of water with BN substrates that vary
in size and dimension, from zero to 2D periodicity. In order to gain a better
understanding about how the interaction varies, we compute reference adsorption
energies, static polarizabilities, and dynamic polarizabilities for borazine
(the BN analogue of benzene), boronene\footnote{We refer to this molecule as
boronene in this study for convenience.} (BN analogue of coronene), and h-BN.
These BN substrates are electronic insulators with band gaps exceeding 4 eV at
the GGA-DFT level, they contain lone pairs of electrons located on the nitrogen
atoms, and they are geometrically analogous to carbon substrates. Intuitively,
the interaction of molecular and surface BN substrates with water is expected to
involve a mixture of polarizable electrostatics and dispersion interactions.
This complex combination of interactions is widespread in biology and surface
science, making the water/BN systems an excellent case-study for establishing
reference adsorption energies. Moreover, like graphene, the 2D h-BN surface also
gives rise to long-range Coulomb interactions that are also relevant to
adsorption on low-dimensional extended systems\cite{ambrosetti2016wavelike}.
However unlike graphene and metallic surfaces, h-BN is an insulator and
therefore it is not well understood whether it gives rise to substantial
non-additive dispersion interactions. Using the static and dynamic
polarizabilities that are computed in this study, this important question on
non-additive dispersion is addressed.

It is worth noting that a series of studies has
ensued\cite{al-hamdani2,wu2015interaction,wu2016hexagonal} since we computed
reference DMC interaction energy curves for water on h-BN\cite{al-hamdani2}. In
our previous work, DMC interaction energy curves were computed for two stable
configurations of water above the h-BN surface and the adsorption energy was
found to be $-85\pm5$ meV at the more favourable site, with one hydrogen atom of
water pointing down towards a nitrogen atom in the h-BN
surface\cite{al-hamdani2}. Furthermore, Wu \etal\ computed the interaction
energy of water on a borazine molecule with DMC, MP2 and
CCSD(T)\cite{wu2015interaction}. Therein, the configuration is such that the two
hydrogen atoms of water point down towards a boron and a nitrogen atom in
borazine\cite{wu2015interaction}. Although the configuration that was used is
unlikely to be the lowest in energy, they showed that CCSD(T) and DMC predicted
the same interaction energy curves, with a maximum interaction of $\sim75$ meV
at 3.36 \AA\ separation\cite{wu2015interaction}. In addition, the RPA method has
been applied to the water/h-BN system with different unit cell
sizes\cite{wu2016hexagonal}, estimating a finite size effect correction of
$-16\pm8$ meV. This correction arises because interactions in the water/h-BN
system can extend beyond the finite unit cell employed, leading to spurious
correlation interactions between electrons and their periodic images.

Here, we used DMC, lattice regularised DMC (LRDMC), CCSD(T), the RPA and MP2
calculations to ascertain the best possible adsorption energy for water on the
molecular and surface BN substrates. These benchmark calculations result in a
particularly important finding: The water adsorption energy remains almost
constant as the BN substrate size and dimension increases.  To explain this
peculiar finding and provide further benchmark data, we perform time-dependent
DFT (TD-DFT) calculations of the static and dynamic polarizabilities for the
water/BN systems. We find that there is considerable non-additivity in the
dispersion interactions of these water/BN systems due to anisotropy. In what
follows, the many-body dispersion (MBD) contribution\cite{mbd,mbd14} in DFT is
found to agree with the RPA and TD-DFT and importantly, it is shown to capture
anisotropic interactions in the long-range limit in these systems. The results
indicate that the MBD scheme is an efficient approach for estimating the
long-range correlation interactions and can be used to determine the finite size
effect correction for explicitly correlated electronic structure methods.
Furthermore, a selection of DFT exchange-correlation (\textit{xc}) functionals
is benchmarked for this set of water/BN substrates. Their performance varies and
we make note of which \xc functionals predict at least the same trend as the
reference methods.

Details of the calculations and set-ups are given first in Section \ref{methods}
followed by results in Section \ref{results}. In Section \ref{energies} the
interaction energy curves for a water monomer with borazine (BN analogue of
benzene, B$_3$N$_3$H$_6$), boronene (BN analogue of coronene,
B$_{12}$N$_{12}$H$_{12}$), and h-BN surface are reported, using a series of
reference electronic structure methods along with previous results from DMC. In
addition, the long-range correlation energy contribution to the water/h-BN
adsorption energy is determined with periodic MP2, the RPA, and MBD, by
computing increasing unit cell sizes of h-BN. This is followed by a brief
assessment of some recently developed and otherwise widely used \xc functionals
in Section \ref{benchmark}. In Section \ref{tddft} we present the static and
dynamic polarizability of water/BN systems from TD-DFT and compare with DFT+MBD.
We close with conclusions in Section \ref{conc}.

\section{Methods}\label{methods} In this study, the absolute interaction
energies of water with BN substrates have been computed using DFT, the RPA, MP2,
CCSD(T), DMC, and LRDMC. The procedure for each of these methods is described
shortly, however let us first define the interaction energy and the systems
being considered in the following section.

\subsection{System setup} The interaction energy between water and the substrate
with all of the electronic structure methods considered here has been calculated
in the same way as in our previous work\cite{al-hamdani2}, \begin{equation}
E_{int}=E^{tot}_{d}-E^{tot}_{\text{far}}\label{BE} \end{equation} where
$E^{tot}_{d}$ is the total energy of water and substrate at a given
oxygen-substrate separation distance, $d$, and $E^{tot}_{\text{far}}$ is the
total energy of water and substrate at 8 {\AA} oxygen-substrate
distance.\footnote{The residual interaction of water with h-BN at 8 {\AA}
distance is less than 4 meV.}
\begin{figure} \centering \includegraphics[width=0.5\textwidth]{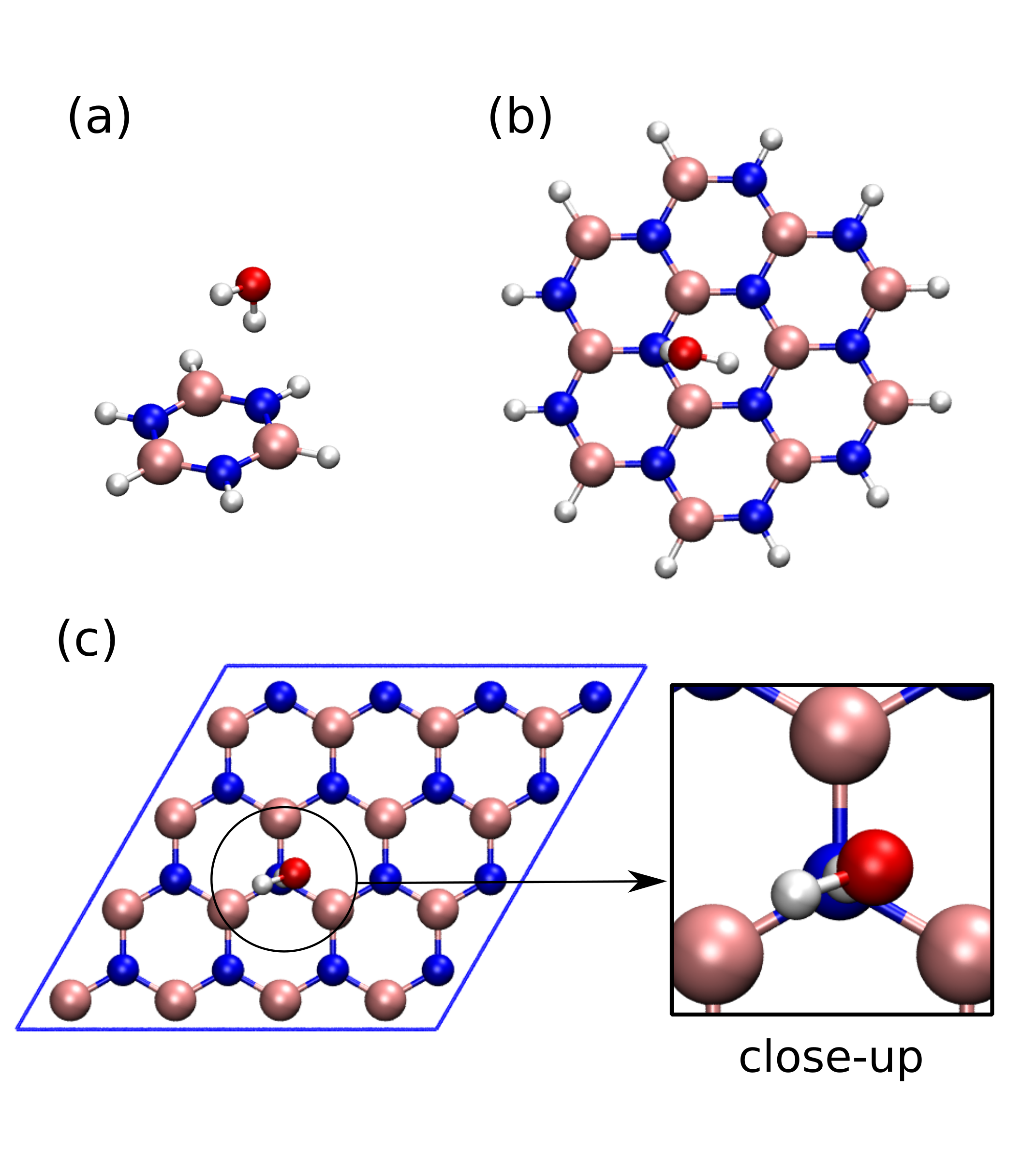}
\caption{(a) Water adsorbed on borazine (B$_3$N$_3$H$_6$). (b) Water adsorbed on
boronene (B$_{12}$N$_{12}$H$_{12}$). (c) Water adsorbed on a (4$\times$4) unit
cell of a h-BN sheet with a close-up of the physisorption site. White spheres
are hydrogen, red spheres are oxygen, pink spheres are boron, and blue spheres
are nitrogen.}\label{figure1} \end{figure}
The water molecule has the same orientation on each substrate, with one hydrogen
atom pointing down towards a nitrogen atom in the substrate. This makes for a
cleaner comparison (see Fig. \ref{figure1}) and in this way, the same low energy
configuration of water/h-BN is used as in Ref. \citenum{al-hamdani2}. DFT
geometry optimizations of the water/BN complexes confirmed that most stable
orientation of the water molecule is the same on these three BN substrates.  The
lattice constant of h-BN used in all periodic calculations is 2.51 \AA, in
agreement with the experimental lattice constant\cite{Bosak2006}. Geometries of
the water/BN complexes in the minimum and far configurations can be found in the
Supporting Information (SI). The interaction energy between water and substrate
is plotted as a function of the perpendicular distance between the oxygen atom
of the water molecule and the flat substrate.

\subsection{DFT calculations} The DFT calculations on the molecular BN systems
have been conducted using the Vienna \textit{Ab-initio} Simulation Package
(VASP.5.4.1) and standard projector augmented wave (PAW)
potentials\cite{PAW_99,Blochl1994} with a 500 eV plane-wave cut-off energy and
$\Gamma$-point sampling of reciprocal space. In VASP and Quantum Espresso,
periodic boundary conditions (PBC) must be set and therefore for periodic
calculations of the molecular water/borazine system, a $15\times15\times15$
\AA$^3$ unit cell was used. With this unit cell, the effect of a dipole
correction is negligible ($<$1 meV) in a GGA calculation with VASP. For the
larger boronene molecule, a $20\times20\times16$ \AA$^3$ unit cell was used.

There are countless \xc functionals that could be assessed and therefore we have
chosen only a few functionals in this study which are either new or widely used
for such systems. Hence, a widely used GGA \xc functional, PBE\cite{PBE}, and a
hybrid functional that contains a fraction of exact exchange,
PBE0\cite{PBE0a,PBE0b}, are assessed. Given the importance of vdW interactions
also, vdW-inclusive functionals are expected to be suitable in such systems and
therefore we also test optB86b-vdW\cite{vdw_opt11,B86} and PBE+D3\cite{D3a,D3b}.
In addition, we have tested the more recently developed strongly constrained and
appropriately normalised (SCAN) functional of Sun \etal\cite{scan} Furthermore,
the interaction energy curves for water/BN substrates have been computed with
PBE+MBD\cite{PBE,mbd,mbd14} and PBE0+MBD\cite{PBE0a,PBE0b,mbd,mbd14} using the
all-electron FHI-AIMS code with tight settings\cite{blum2009ab}.

\subsection{Molecular quantum chemistry calculations} The molecular systems were
also calculated with quantum chemical codes (Gaussian \cite{g03}, Molpro
\cite{MOLPRO_brief}, and NWChem \cite{valiev2010nwchem}) to obtain the MP2
interaction energy at 3.4 \AA\ water/substrate distance, with complete basis set
(CBS) extrapolation\cite{extrapolation} using Dunning's aug-cc-pV(T,Q)Z basis
sets\cite{dunning1989,feller1996}.  In order to compute interaction energies
with reference accuracy, we use the linear scaling domain based pair natural
orbital CCSD(T) (denoted here as LCCSD(T)) method\cite{dlpno} as implemented in
the ORCA program package.\cite{orca} The implementation has been optimized to
use compact representations of all amplitudes and imposing block sparsity of
tensors.\cite{dlpno_compact} The pair natural orbitals are used to employ the
locality of electron correlation resulting in tunable thresholds to approach the
full CCSD(T) `gold standard' of quantum chemistry. The domain and pair
thresholds are used with the tight settings as necessary for the accurate
computation of non-covalent interactions. Tightening the thresholds further for
water/borazine led to a change in the interaction energy of less than 5 meV.


\subsection{Periodic quantum chemistry calculations}  A developer version of
VASP 5.4.2 was used for the periodic MP2~\cite{marsman2009second} and the
RPA+SOSEX~\cite{grueneis2009making,kaltak2014cubic,kaltak2014low,klimevs2015singles}
calculations for water on h-BN within the PAW framework. In all calculations the
electronic states of the H atoms were treated as valence states while the $1s^2$
states of B, N and O atoms where kept frozen. The $\Gamma$-point approximation
was used to sample the first Brillouin zone. MP2 calculations utilize HF
canonical orbitals, while the RPA caclulations employ PBE orbitals. For the MP2
and RPA methods the one-electron states were expanded using a 600 eV plane-wave
energy cut-off. We employ a set of atom-centered functions to construct the
unoccupied one-electron states, mapped onto a plane-wave representation.
Dunning's contracted aug-cc-pVDZ (AVDZ) and aug-cc-pVTZ (AVTZ) basis
sets~\cite{dunning1989,feller1996} were used as atom-centered functions. With
the rediagonalization of the Fock matrix in this newly constructed virtual space
we mimic an AVDZ and AVTZ basis set in a plane-wave basis suitable for periodic
correlated calculations~\cite{booth2016}. Counterpoise corrections (CP) to the
basis set superposition error (BSSE) were included in all correlated
quantum-chemical calculations which involve pseudized Gaussian type orbitals
(PGTOs).  If not stated otherwise, we extrapolate the correlation energies to
the complete basis limit using AV(D,T)Z results. The two-electron integrals are
calculated using an auxiliary plane-wave basis in
VASP~\cite{marsman2009second,grueneis2010mp2,grueneis2015explicitly}. The
kinetic energy cutoff $E_{\chi}$, such that $\frac{\hbar^2 \mathbf{G}^2}{2m_e} <
E_{\chi}$ defining this auxiliary plane-wave basis set was set to 300~eV. All
reported MP2 and RPA adsorption energies have been checked for convergence with
respect to this cutoff. We note that the canonical implementation of MP2 in {\sc
VASP} scales as $\mathcal{O}(N^5)$~\cite{marsman2009second} although recently an
implementation scaling as $O(N^4)$ became available as
well\cite{schafer2017quartic}. The RPA calculations with PGTOs were also
performed in an $\mathcal{O}(N^5)$ implementation in VASP allowing for the
inclusion of second-order screened exchange (RPA+SOSEX) for a more accurate
description~\cite{grueneis2009making}.

Additional RPA calculations were preformed using a cubic scaling implementation
in VASP~\cite{kaltak2014cubic,kaltak2014low,klimevs2015singles} and the full
plane-wave basis in contrast to the PGTOs. PBE orbitals were
used\footnote{Differently to current default settings in VASP, NOMEGA = 8 was
set and PRECFOCK = Normal for the RPA calculations.} along with a 430 eV
plane-wave energy cutoff and the $\Gamma$-point approximation to sample the
first Brillouin zone. In this case the cutoff for the response functions was set
to the default value of 287 eV. The results were extrapolated to the infinite
basis set limit using the internal VASP extrapolation. This assumes that errors
drop off like one over the basis set
size\cite{harl2008cohesive,klimevs2014predictive}. A quadrature with 8 grid
points was used for the evaluation of the imaginary time and frequency
integrations. In addition, the contribution from GW single excitations (GWSE)
were computed based on the work of Klime{\v{s} \etal\cite{klimevs2015singles} We
stress that adsorption energies calculated within the RPA using the cubic
scaling implementation and the $\mathcal{O}(N^5)$ implementation with PGTOs are
in good agreement.\footnote{For a ($4\times4$) h-BN unit cell, the cubic scaling
implementation predicts adsorption energies of $-65$ meV, whereas the basis set
extrapolated PGTO value is $-69$ meV. Increasing the cutoff from 430 eV to 500
eV in the cubic scaling code does not change these values. The slight
differences are most likely related to different basis set extrapolation
procedures.}

\subsection{Fixed-Node Quantum Monte Carlo}
Herein we report results coming from two different fixed-node quantum Monte
Carlo (FNQMC) approaches: standard DMC\cite{foulkes01} and
LRDMC\cite{LRDMC:prl2005, casula10}.  These are both projection Monte Carlo
methods: they can access the electronic ground state energy of the system by
iteratively projecting an initial trial wave function $\psi_T$ into the ground
state, with the constraint that the projected wave function $\Phi$ has the same
nodal surface of an appropriately chosen guiding function $\psi_G$ (fixed node
approximation).\cite{Reynolds:1982en,foulkes01} Typically,
$\psi_T=\psi_G=\psi_\text{VMC}$, where $\psi_\text{VMC}$ is the best function
obtained within a variational Monte Carlo approach. Whenever $\psi_G$ has the
exact nodal surface, the approach is exact, otherwise it gives the best
approximation of the ground state given the fixed node constraint.

In projection Monte Carlo approaches there is a second approximation
in how the projection is performed, and it is different in DMC and
LRDMC.  The projection in DMC comes from the imaginary time
Scr\"{o}dinger equation; it is implemented as an imaginary time
evolution, where a time-step $\tau$ has to be chosen.  The chosen
$\tau$ is a trade-off between accuracy and computational cost: the
latter is $\propto 1/\tau$, but the projection is exact only in the
limit of $\tau \to 0$.
Recently, an improved DMC algorithm\cite{zen2016boosting} that solves a
size-inconsistency issue at finite values of $\tau$ was introduced, and it was
shown that it dramatically reduces the time-step errors in the evaluation of
interaction energies.

Here, DMC calculations were performed with the CASINO code\cite{casino}, using
Slater-Jastrow type trial wavefunctions in which the Jastrow factor contains
electron-nucleus, electron-electron, and electron-electron-nucleus terms. We
used Trail and Needs pseudopotentials\cite{TN1,TN2} for all atoms. The DMC
procedure is similar to that used in Refs. \citenum{al-hamdani1,al-hamdani2}.
The initial single particle wavefunctions for use in DMC were obtained from DFT
plane-wave calculations using Quantum Espresso v.5.0.3 \cite{pwscf}. A standard
300 Ry energy cutoff was applied and for efficiency, the resulting wavefunctions
were expanded in terms of B-splines\cite{bsplines} using a grid multiplicity of
2.0 corresponding to a grid spacing, $a = \pi/2G_{max}$, where $G_{max}$ is the
plane wave cutoff wavevector. Trial wavefunctions were generated using the
LDA\cite{LDA} which has been validated for weak interactions in previous
work\cite{Ma_11a}. VMC was used to optimize the Slater-Jastrow type trial
wavefunctions. In the evaluation of weak interactions in systems such as those
under consideration in this work, the standard Jastrow correlated single Slater
determinant (Slater-Jastrow) has proved to be sufficiently accurate, as
established in a number of studies\cite{Ice:prl2011, santra_on_2013,
Quigley:Ice_0_i_Ih:jcp2014, Morales:bulkwat:2014, Cox:2014, Benali:2014,
al-hamdani1, al-hamdani2, gillan15, Morales:perspective2014, noncov:chemrev2016,
Zen-liquidwat, Ice2D_2016, KaoliniteQMC}.

In DMC, the locality approximation was utilized \cite{locapp} with a time-step,
$\tau_{standard}$, of 0.015 a.u. which was tested against a time-step,
$\tau_{small}$, of 0.005 a.u for the water/borazine interaction. The interaction
energy agrees within the error bars of less than 5 meV for $\tau_{standard}$ and
$\tau_{small}$.\footnote{Note that in the time this work has been conducted, Zen
\etal~ implemented a more efficient method in the CASINO code that allows for
better convergence with large time-steps\cite{zen2016boosting}. We tested the
water/boronene system with this new implementation for time-steps of 0.050 a.u.
(considered as $\tau_{standard}$ with the new implementation) and 0.025 a.u.
(considered as $\tau_{small}$). The water/boronene interactions are converged
within the stochastic error for both implementations and time-steps in CASINO.}

LRDMC, on the other hand, is based on the spatial discretization of the
molecular Hamiltonian on a lattice of mesh size $a$, and it resorts to the
projection scheme used also in the Green function Monte Carlo
algorithm.\cite{Sorella:2000p17651,Buonaura:1998p25304} The error induced by the
finite mesh size $a$ is analogous to the time step error appearing in standard
DMC calculations; it can be controlled by performing several calculations with
different values of the mesh $a$ and finally extrapolating to the continuum
limit $a\to0$, but in practical cases it is sufficient to consider the results
for a mesh $a$ small enough that the expected finite-mesh bias is
negligible.\footnote{LRDMC\cite{LRDMC:prl2005, casula10} preserves the
variational principle even when used in combination with nonlocal
pseudopotentials, and it is size-consistent for any value of the mesh $a$,
maintaining its efficiency even for systems with a large number of electrons. }

The LRDMC results reported in this work have been obtained with the {\sc
TurboRVB} package developed by Sorella and coworkers.\cite{TurboRVB} We used a
Slater-Jastrow trial wavefunction, but the setup for the LRDMC calculations
slightly differs from that of the DMC calculations as a consequence of the
different implementations of the algorithms in {\sc TurboRVB} and {\sc CASINO}.

For a more detailed description of the functional form of the wavefunction
implemented in {\sc TurboRVB} see Ref.  \citenum{Zen:2013is}. Core electrons of
B, N, and O atoms\footnote{ We used a $(7s,2p)$ basis for the H atoms,
$(9s,10p,2d)$ for B, $(10s,10p,2d)$ for N, $(10s,11p,2d)$ for O, all of them
with Gaussian-type atomic orbitals. Note that, due to the presence of the
Jastrow factor, the orbitals with too high value for the exponent or with very
high angular momentum, lead to an almost negligible improvement in the
variational wave function quality, in front of a much slower and inefficient
optimisation of the parameters.\cite{Zen:2013is} Thus, the B, N and O exponents
have been taken starting from the Burkatzki {\em et al.} VTZ basis
set\cite{filippi-basis}, that we have uncontracted and removed the $f$ orbitals,
the largest exponents ({\em i.e.}, $> 30$) or the almost redundant ones. For H,
which has no pseudopotential, we used an all-electrons basis set.} have been
described via scalar-relativistic energy-consistent Hartree-Fock (HF)
pseudopotentials of Burkatzki {\em et al.}\cite{Burkatzki:2007p25447}
The coefficients of the molecular orbitals have then been optimized by
performing an LDA calculation, using the DFT code included in the {\sc TurboRVB}
package.\cite{TurboPREP} The Jastrow factor used here consists of terms that
account for the electron-electron, electron-nucleus and
electron-electron-nucleus interactions.\footnote{The non-homogeneous terms are expressed
in terms of atomic orbitals, which are expanded in terms of a $(2s,2p)$ basis
for H atoms, $(3s,2p,1d)$ for B, N and O atoms.}  The exponents of the Jastrow
atomic orbitals have been fixed to the values obtained from the optimisation in
the water/borazine system.  All the other parameters of the Jastrow factor have
been optimized for each specific configuration.  In the LRDMC we used a mesh $a$
of 0.3~a.u.
We verified in the water/borazine system that we have no bias given by
the choice of the pseudopotentials (indeed, by using the Trail and
Needs pseudopotentials\cite{TN1,TN2} we obtained results in agreement
with those from Burkatzki {\em et al.}, within the statistical error
margin), or by the choice of the mesh $a$.  This is not a surprise,
since the approach here adopted was already tested in a number of
other systems.\cite{Zen:2013is, Zen2014:ROS, ZenDir:JCTC2014,
  ZenRet:JCTC2015, Zen-liquidwat, KaoliniteQMC}
%

Energies from quantum Monte Carlo approaches -- both DMC and LRDMC -- are
affected by finite size simulation errors,\cite{FSEqmc:PRB2008,
FSE:Ceperley2016} which in the h-BN system under consideration here is mostly
arising from the long-range Coulomb interactions. There are currently three
methods available to deal with them: the use of the model periodic Coulomb (MPC)
potential for the long-range
interaction\cite{MPC:Fraser1996,MPC:Will1997,MPC:Kent1999}, or {\em a
posteriori} corrections to the energy as prescribed by
~\citet{Chiesa:size_effects:prl2006} or Kwee, Zhang and Krakauer
(KZK)\cite{KZK:prl2008}. In this work we have used MPC, one of the most accurate
approaches\cite{azadi2015jcp}.

In all systems, the interaction energy was evaluated as the difference between
the bound configuration and the configuration with the water molecule far away,
according to Eq.~(\ref{BE}).  This choice leads to an efficient cancellation of
errors due to the finite time-step (for DMC)\cite{zen2016boosting} or mesh (for
LRDMC), and finite-size simulation\cite{KaoliniteQMC}.
For the water/borazine and water/boronene systems we used open boundary
conditions in FNQMC calculations.  For the water/h-BN system we used 3D periodic
boundary conditions, having set the distance between h-BN sheets to 16~\AA.

\section{Results}\label{results} Reference interaction energy curves for water
on borazine, boronene and h-BN at the most favourable adsorption site have been
computed, contributing to the body of knowledge that has been reported
previously\cite{wu2015interaction,al-hamdani2}. The long-range correction to the
adsorption energy in the water/h-BN system is determined from FNQMC, the RPA, MP2,
and MBD at the most favourable adsorption site. In what follows, it will be
shown that excellent agreement is achieved amongst the benchmark methods and we
provide our best prediction of the water/h-BN adsorption energy. We also compare
some \xc functionals and find that their performance varies across the
dimensions. Later we present TD-DFT results of the static and dynamic
polarizabilities for these water/BN systems and using these properties we
explain the trend observed for the water adsorption energy on BN substrates from
zero to 2D. In doing so we also demonstrate the accuracy of the MBD method in
DFT.

\subsection{Molecular adsorption energies on BN substrates from zero to 2D}\label{energies}

In order to predict and understand how the water/BN interaction properties vary
from a small borazine molecule up to the extended 2D h-BN surface, accurate
prediction of the adsorption energies is necessary. To this end, reference DMC
interaction energy curves have been computed here for water on borazine and
boronene and are shown in Fig. \ref{figure2}. We are referring to the BN
analogue of coronene as boronene here, also used by Wu \etal\ in an
extrapolation scheme to predict the interaction energy of water on
h-BN\cite{wu2015interaction}. In addition, Al-Hamdani \etal\ computed the
interaction energy curve for water at a boron and nitrogen site in h-BN from DMC
(the latter is also shown in Fig. \ref{figure2}). Later, Wu \etal\ performed
direct RPA calculations and estimated a $16\pm8$ meV finite size error (FSE)
correction\cite{wu2016hexagonal}. The finite size correction is necessary
because of the long-range Coulomb interactions between fluctuating charges that
are exhibited by this system. Here, we first demonstrate the effect of single
excitations (SE) and second order screened exchange (SOSEX) on the RPA
interaction energies and second, we find the extent of the long-range charge
fluctuations and therefore the finite size correction with periodic MP2, the RPA
and the MBD method in DFT. In addition, we also compute the FSE in FNQMC
interaction energies from the KZK and MPC methods.

The DMC interaction energy curves for water/borazine and water/boronene have
been computed here using the configurations of water shown in Fig. \ref{figure1}
and a DMC time-step of 0.015 a.u. For water/borazine the DMC interaction energy
minimum appears at 3.32 \AA\ in Fig. \ref{figure2}a. Further reference
interaction energies from LCCSD(T), LRDMC, DMC (with 0.005 a.u. time-step), MP2,
and the RPA are reported for water/borazine at this 3.32 \AA\ oxygen-ring
distance in Table \ref{table1}. The LRDMC, LCCSD(T) and MP2 results agree with
the DMC interaction energy within the stochastic error bars that are $<5$
meV.\footnote{It should be noted that without CBS extrapolation, there is a 20
meV overestimation for the MP2 energy with Dunning's aug-cc-pVTZ basis set, and
it is therefore essential to employ basis set extrapolations.} Note that we have
also computed the conventional CCSD(T) interaction energy for the water/borazine
system and at $-119$ meV, it is in close agreement with LCCSD(T).  The direct
RPA is instead underestimating the water/borazine interaction by $\sim20$ meV.
Indeed, the direct RPA has been found to underbind weakly interacting systems in
general\cite{ren2011beyond}. This underestimation can be alleviated by including
GWSE correction and for water/borazine this leads to an $20$ meV improvement,
bringing the RPA+GWSE into agreement with MP2, DMC, LRDMC, and LCCSD(T). The
GWSE contribution is determined by calculating the first order difference of the
density matrix between DFT and the GW approximation and correspondingly, a first
order correction to the HF energy functional\cite{klimevs2015singles}. This
approach yields a slightly smaller correction than the single excitations
correction of Ren and coworkers\cite{ren2011beyond}. In the GWSE approximation,
correlation effects in the density matrix are more accurately included. In the
present case, the standard SE contributions are about 25 meV whereas, the GWSE
contributions amount to the already quoted 20 meV.

On the larger boronene molecule, the water interaction minimum appears at 3.40
\AA\ on the DMC interaction energy curve shown in Fig. \ref{figure2}b. The main
difference between boronene and borazine is their size and hence, the water
adsorption energy is naively expected to be larger on boronene than on the
smaller borazine molecule due to dispersion. However, using the 3.40 \AA\
oxygen-boronene separation distance in Table \ref{table1}, the water/boronene
LCCSD(T) interaction energy is $-109$ meV which is $\sim10$ meV less than the
water/borazine interaction.  MP2 is also in close agreement with LCCSD(T) for
water/boronene. However, the LCCSD results show that the contribution from
perturbative triple excitations is a considerable 20 meV for the water/borazine
and water/boronene systems.  Therefore, the performance of MP2 appears to be
fortuitous. This has important implications for extrapolation schemes, where it
is not given that the fortuitous behavior in MP2 is systematic. As seen for
water/borazine, LRDMC agrees with LCCSD(T), predicting an interaction energy of
$-107\pm5$ meV for water/boronene.  We also see an indication that DMC predicts
a slightly smaller interaction energy than LCCSD(T) and LRDMC. The difference is
small considering the stochastic error on the DMC energy, but given the slightly
better agreement between LRDMC and LCCSD(T) we suggest LRDMC provides the best
prediction for water adsorption on h-BN in the following.
\begin{table}[tbp] \begin{threeparttable} \begin{ruledtabular}\centering
\caption{Interaction energies using different methods for water with borazine
(at 3.32 \AA\ oxygen height above ring), boronene (at 3.40 \AA\ oxygen height
from the molecule), and h-BN. Where possible, the oxygen/h-BN distance at the
minimum of the interaction energy curve with the corresponding method is shown
in parenthesis in \AA. Canonical coupled cluster theory has been used for
water/borazine whilst local coupled cluster theory has been used for
water/boronene. The water/h-BN adsorption energies include a finite size
correction of 20 meV, except for the FNQMC results which are corrected with the
MPC method (10 meV). Converged DFT results are reported in the bottom panel.}
\label{table1}
      \begin{tabular}{lrrr}
        Method                & Water/Borazine & Water/Boronene  & Water/h-BN   \\
        (L)CCSD(T)/CBS        &$ -123$         & $-109$          & n/a          \\
        LRDMC                 &$ -122\pm4$     & $-107\pm5$      & $-107$$\pm7$ (3.40)  \\
        DMC                   &$ -117\pm3$     & $-96\pm8$       & $ -95$$\pm5$ (3.40)  \\
        MP2/CBS               &$ -120$         & $-113$          & $-110$ (3.25)\\
        (L)CCSD/CBS           &$ -102 $        & $-90$           & n/a          \\
        RPA+SOSEX+GWSE        &  n/a           & n/a             & $-113$ (3.25)\\
        RPA+GWSE              &$ -112 $        & n/a             & $-108$ (3.25)\\
        RPA                   &$ -92$          & n/a             & $ -89$ (3.36)\\ \hline
        PBE                   &$ -82$   &$-46$   &$-44 $ (3.40)   \\
        SCAN                  &$-122$   &$-103$  &$-99 $ (3.20)   \\
        PBE+D3                &$-131$   &$-132$  &$-130$ (3.20)   \\
        PBE0+MBD              &$-134$   &$-130$  &$-135$ (3.20)   \\
        optB86b-vdW           &$-125$   &$-151$  &$-168$ (3.20)   \\
        vdW-DF2               &$-118$   &$-129$  &$-141$ (3.30)
      \end{tabular}
    \end{ruledtabular}
   \end{threeparttable}
\end{table}
\begin{figure}[ht]
\centering
\includegraphics[width=1.00\textwidth]{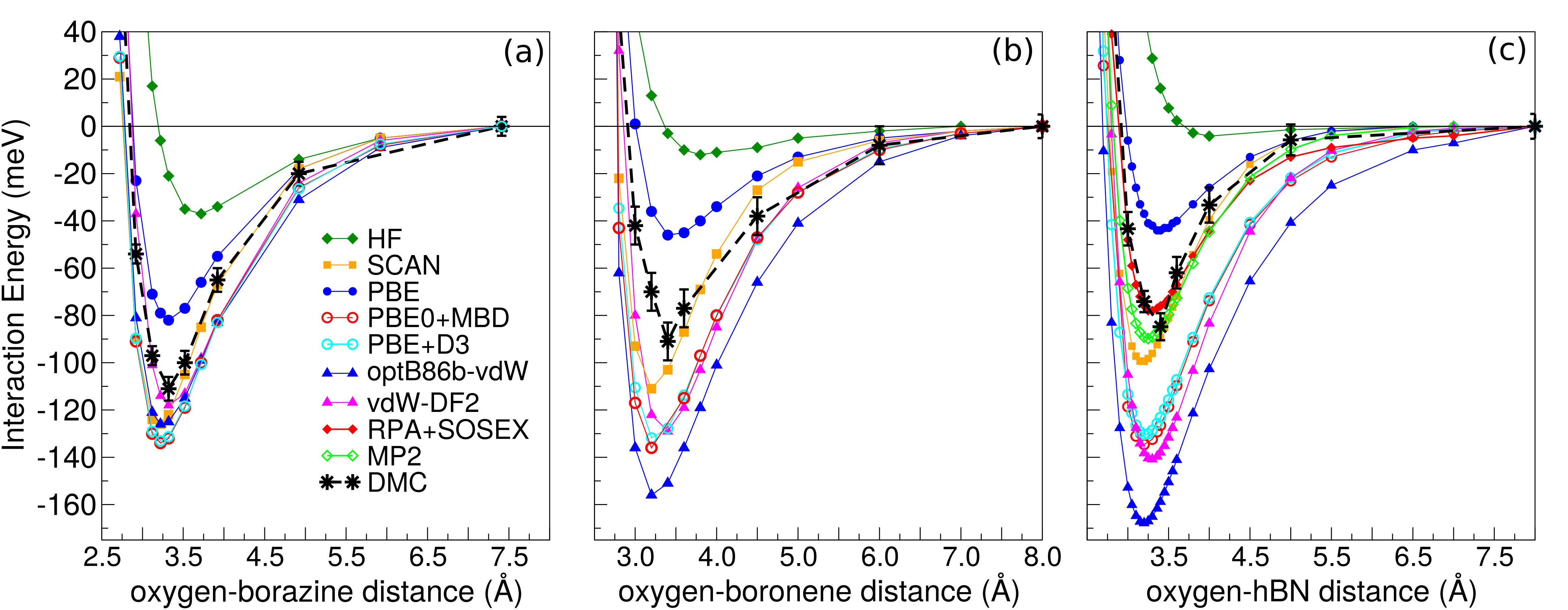}
\caption{Interaction energy curves for water situated above the N site
  in: (a) borazine molecule (B$_3$N$_3$H$_6$), (b) boronene molecule
  (B$_{12}$N$_{12}$H$_{12}$) and (c) h-BN surface. The DMC data points
  for water/h-BN are taken from a previous study, see
  Ref. \protect\citenum{al-hamdani2}. The DMC data shown has been
  computed with $\tau_{standard}=0.015$~a.u. The data points are connected by
  lines simply to guide the eye. }\label{figure2}
\end{figure}

Interaction energy curves are also shown for water/h-BN from DMC, the RPA, the
RPA+SOSEX and MP2 (see Fig. \ref{figure2}c). MP2 calculations are performed for
the periodic system, with the same unit cell as for the other methods. The RPA,
RPA+SOSEX, and MP2 energies have been extrapolated to the complete basis set
limit using pseudized aug-cc-pV(D,T)Z basis sets. Note that the interaction
energy curves in Fig. \ref{figure2}c are computed for the $(4\times4)$ unit cell
of h-BN and the values reported now are taken directly from these curves without
including a correction for finite size effects. It can be seen from Fig.
\ref{figure2}c that the RPA interaction energies agree within the stochastic
error of DMC interaction energies - in line with the findings of Wu
\etal\cite{wu2016hexagonal} The interaction energy of water according to the RPA
is $-69$ meV at the minimum (3.36 \AA). From Fig. \ref{figure2} the SOSEX
correction to direct RPA increases the interaction energy on the order of
$\sim5$ meV, especially when water is close to the h-BN surface at $<$3.5 \AA\
height. Therefore, the RPA+SOSEX interaction energy of water on h-BN is $-74$
meV at the minimum (3.25 \AA) on the RPA+SOSEX interaction energy curve. Earlier
however, GWSE was found to contribute 20 meV towards the interaction energy of
water/borazine and similarly here, GWSE contributes 20 meV to the interaction
energy of water on h-BN at the minimum. As such, the RPA+SOSEX+GWSE adsorption
energy of water/h-BN is $-94$ meV. Meanwhile, MP2 predicts an adsorption energy
of $-88$ meV for water/h-BN, also at 3.25 \AA. It should be noted that MP2
theory does not contain singles contributions because of Brillouin's theorem.
The interaction energies from RPA+SOSEX+GWSE and MP2 are slightly larger than
the DMC interaction energy although, these are still within a few meV of the DMC
adsorption energy ($-85\pm5$ meV). Furthermore, the LRDMC interaction energy of
water on h-BN at 3.40 \AA\ is $-97\pm7$ meV. Thus all of the aforementioned
reference methods are demonstrably in excellent agreement. However, as
previously indicated by Wu \etal\cite{wu2016hexagonal}, there is a long-range
Coulomb interaction that arises in methods that account for fluctuating dipoles
which is not fully captured in the ($4\times4$) unit cell of h-BN at the
$\Gamma$-point. Since this interaction extends beyond the $(4\times4)$ unit cell
of h-BN, it requires a great deal of computational effort to capture with
expensive many-body wavefunction based methods. As such, the agreement between
different methods has scarcely been verified on an extended surface. Let us
address this particular point in the following.

Long-range Coulomb type interactions from the fluctuation of electrons,
otherwise referred to as long-range correlation energy, can extend to the
nanometre scale\cite{ambrosetti2016wavelike} and are expected to be particularly
important in low dimensional systems. In methods that account for charge
fluctuations (and therefore vdW interactions) explicitly, whether it is by $s
\rightarrow p$ excitations for coupled harmonic oscillators~\cite{Hermann2017}
or via more general excitation mechanisms in the RPA and MP2 theory, the
fluctuations can give rise to spurious interactions with their periodic images.
As such, correlated methods are affected by FSE any time they are providing the
energy of a macroscopic system by employing PBC. Since the interaction energy is
evaluated as the difference between two periodic systems here, a big, but only
partial, cancellation of the FSE can be expected. These FSE are much larger than
those observed in (effectively) independent electron methods such as HF and DFT.

To establish the extent of FSE in methods that account for correlation effects explicitly,
MP2,\footnote{After checking convergence, a 500
eV energy cutoff was employed for the one-particle HF states along with
$\Gamma$-point sampling of the Brillouin zone. The cutoff energy for the
auxiliary plane-wave basis set required for the evaluation of the two-electron
four-orbital integrals\cite{marsman2009second} was set to 250 eV. Occupied HF
states were converged within the full plane-wave basis, whereas the virtual
orbitals were constructed using Dunning's contracted aug-cc-pVDZ and aug-cc-pVTZ
\cite{dunning1989,feller1996} pseudized Gaussians in a plane-wave
representation, projected to the HF occupied states.}
RPA, RPA+GWSE and
PBE+MBD interaction energies for water/h-BN have been computed in increasing
supercell sizes. More specifically, h-BN systems with as many as 98 surface
atoms in the unit cell have been computed with MP2 and the RPA in order to
establish the long-range behavior of the water interaction energy.  MP2
calculations have been performed using a pseudized AVTZ basis set, whereas RPA
calculations utilize the cubic scaling implementation in VASP.  Note that this
is a factor of 2 larger than the typical unit cells used in molecular adsorption
studies on low dimensional
surfaces\cite{Voloshina_11,al-hamdani2,Ma_11b,binnie2011ab,wu2015interaction}.
This is a brute force approach, computationally very demanding because the
scaling with size of the correlated methods and as such, the FSE is computed
only near the minimum of the interaction energy curve. The results are reported
in Table \ref{table2} and shown in Fig. \ref{figure3}a.  The finite size effect
correction to the interaction energy of water in a $(4\times4)$ unit cell of
h-BN is denoted here as $\Delta E_{int}^{fse}$. Importantly, the various
reference methods we have used, namely, MP2, the RPA, and the RPA+GWSE, as well
as the MBD method in DFT, all estimate $\Delta E_{int}^{fse} \approx 20$ meV
(see Table \ref{table2}) when $\Gamma$-point calculations are performed.
In addition to establishing the agreement amongst
these methods, this 20 meV is approximately $25\%$ of the reference adsorption
energy computed in a $(4\times4)$ unit cell of h-BN and is therefore a
considerable contribution.
It is worth noting in Fig.~\ref{figure3}a the different convergence of DFT
calculations with the PBE functional, which represents a typical case for a
method with effectively independent electrons.  The PBE interaction energy of
water/h-BN is almost converged in a (3$\times$3) unit cell of h-BN. What about
the convergence of vdW-DFs or dispersion corrections such as the D3?
Fig.~\ref{figure3}a shows that in these methods the convergence is similar to
that of PBE, \textit{i.e.} the dispersion interaction given by these methods
is fully captured in the (4$\times$4) unit cell of h-BN.
This behavior is expected since vdW-DFs and dispersion corrections evaluate
dispersion interactions indirectly, from the density of electrons in the former
and from the environment dependent isotropic atomic C$_6$ dispersion
coefficients in the latter -- and not explicitly as in the RPA, MP2 or MBD.
Furthermore, in the D3 method the interactions are not only calculated in a
minimum image convention, but the long-range interactions are summed over
repeated images up to very large distances that exceed the actual simulation
cell.
The results shown in Fig. \ref{figure3}a signify that the MBD method converges
in the same manner as the RPA and MP2 with unit cell size when using the
$\Gamma$-point only.
Therefore, the MBD method can be used as an effective and efficient
estimate of the FSE for other more expensive, explicitly correlated methods.
\begin{table}[tbp]
\begin{threeparttable}
\begin{ruledtabular}\centering
\caption{Interaction energies in meV for water on h-BN at 3.25 \AA\ calculated
using $\Gamma$-point sampling only with increasing supercell sizes. N is the
number of atoms in the h-BN substrate unit cell. $\Delta E_{int}^{fse}$ is the
difference in the interaction energy due to long-range Coulomb interactions for
water with 32 and 98 atoms in the h-BN unit cell.} \label{table2}
\begin{tabular}{lcccc}
N & MP2 & RPA & RPA+GWSE & PBE0+MBD \\ \hline
18 ($3\times3$)        & -70  & -34 & -58     & -88  \\
32 ($4\times4$)        & -93  & -64 & -84     & -125 \\
50 ($5\times5$)        & -106 & -76 & -95     & -139 \\
72 ($6\times6$)        & -113 & -78 & -97     & -145 \\
98 ($7\times7$)        & -116 & -84 & -103    & -148 \\ \hline
$\Delta E_{int}^{fse}$ & -23  & -20 & -19     & -23
\end{tabular} \end{ruledtabular}
\end{threeparttable} \end{table}
\begin{figure}[ht] \centering
\includegraphics[width=1.0\textwidth]{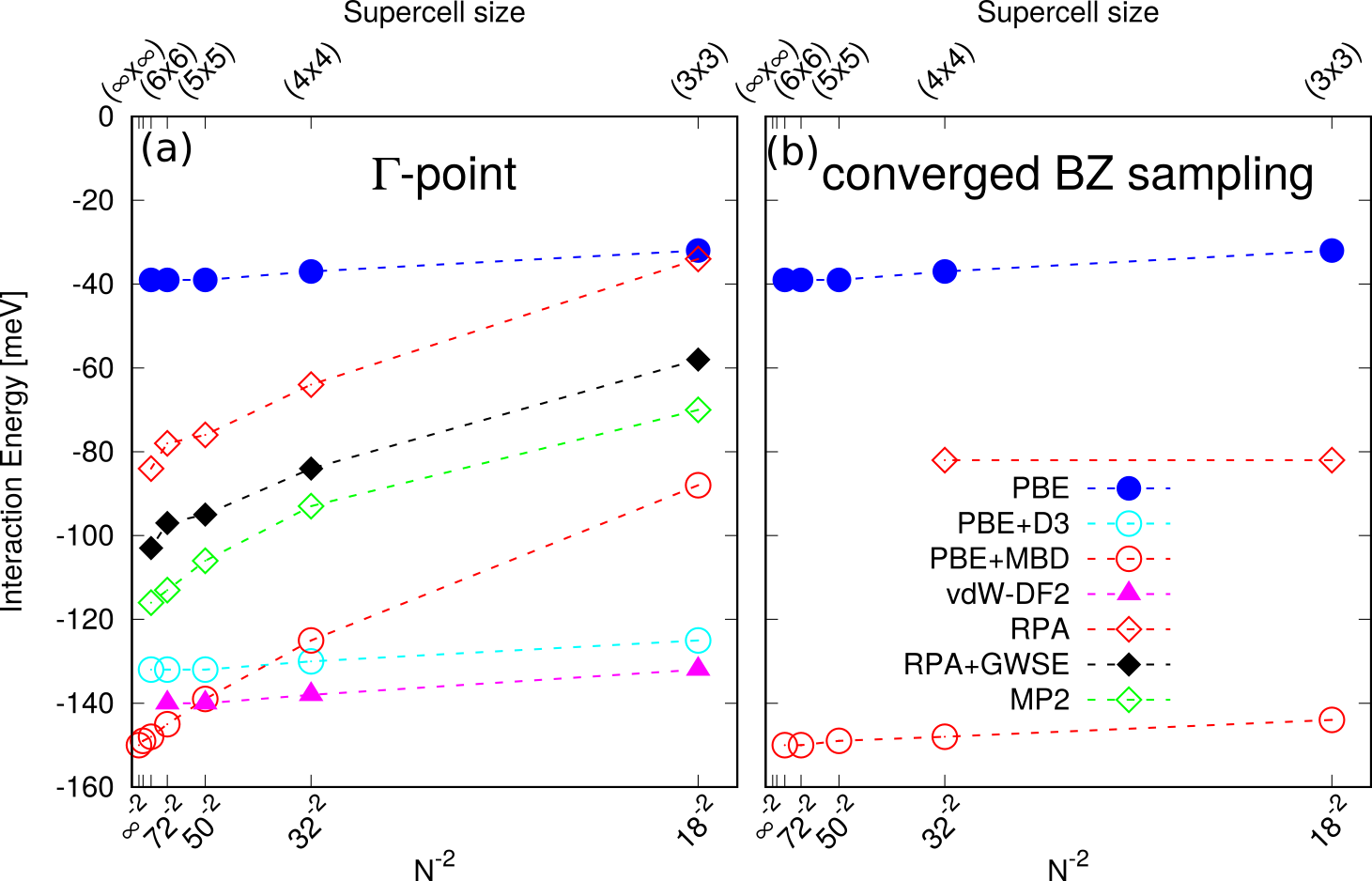} \caption{Interaction
energy of water/h-BN at 3.25 \AA\ height for different supercells with N being
the number of atoms in the h-BN unit cell. (a) Interaction energies with
$\Gamma$ point sampling only. (b) Interaction energies with converged
\textbf{k}-point sampling of the Brillouin zone (BZ). The x-axis is scaled as
1/N$^{2}$ which is the expected scaling from pairwise additive interaction. The
interaction energies are shown for RPA, RPA+GWSE, MP2, PBE+MBD, PBE, PBE+D3, and
vdW-DF2.} \label{figure3} \end{figure}

In methods that account for correlation effects explicitly, a faster cell size
convergence can be obtained by improving the \textbf{k}-point sampling. In the
RPA, using two \textbf{k}-points in the Brillouin zone with the coordinates
(0,0,0) and ($\tfrac{1}{3}$,$\tfrac{1}{3}$,0) yields an adsorption energy of
$-82$~meV instead of $-64$ meV for the ($4\times4$) cell, which is very close to
the low coverage value of $-84$~meV in Table \ref{table2}. Sampling at the
$\Gamma$-point of the ($4\times4$) unit cell of h-BN evidently does not describe
the long-range correlation effects very accurately within these methods.
Likewise, most implementations of the MBD method include very long-range
dipole-dipole interactions by either calculating the MBD interactions in a
supercell of the actually used cell, or by explicitly using a denser
\textbf{k}-point sampling than that which is used in the underlying DFT
electronic structure calculations. Thus, in most cases, the 1/N$^2$ convergence
observed in Fig. \ref{figure3}a can be overcome resulting in a converged
behavior as shown in Fig. \ref{figure3}b.

Fortunately, much cheaper methods for the evaluation of the FSE in correlated
methods exist, and they are only slightly less accurate than the brute force
method. In the FNQMC community there are several schemes such as the KZK, MPC, and
Chiesa methods. We have computed the MPC correction in FNQMC and find it to be 10
meV. Since this correction is based on the FNQMC calculations, we use this value
to correct both the LRDMC and DMC adsorption energies of water on h-BN reported
in Table \ref{table1}. The DFT based KZK correction to FNQMC has also been
computed out of interest and it results in  $\Delta E_{int}^{fse}=17$ meV.
Although the KZK method is not as accurate the MPC method, it predicts a
reasonable estimate of the FSE given that it is less computationally demanding.
Thus the best estimate of the water/h-BN adsorption energy with the MPC
correction from DMC is $-94\pm5$ meV and $-107\pm7$ meV from LRDMC. Recently,
corrections for the correlated quantum chemistry methods have also been
introduced~\cite{liao2016}, and such schemes yield approximately a 20~meV
correction for water adsorption on a $(4\times4)$ unit cell of h-BN for MP2 and
the RPA~\cite{gruber2017}.

To briefly summarise the results of this section, water/BN interaction energies
have been reported from a wide range of electronic structure methods. Given that
methods like LCCSD(T), LRDMC, DMC, MP2 and the RPA are often regarded as
reference methods, it is important to demonstrate that these different
electronic structure theories predict the same interaction energies for
realistic surface models. With a great deal of computational effort that has
been expended here, agreement between the aforementioned explicitly correlated
methods has been demonstrated for water adsorption on molecular and extended BN
substrates. All of these electronic structure methods involve practical
approximations that have had to be carefully addressed in order to predict
accurate adsorption energies. The series of calculations in this study lead to
an improved prediction of the water/h-BN adsorption energy, that is $-107\pm7$
meV with LRDMC. This carefully established water adsorption energy on h-BN
presents a challenge for experiments, and hence it would be particularly
exciting to see future experimental adsorption studies focusing on this system.
In addition, the reference information in this study is intended to help the
development of computational methods and in the following section, we benchmark
a selection of \xc functionals from DFT.

\subsection{Benchmarking \xc functionals in DFT}\label{benchmark}  Let us focus
now on the performance of DFT \xc functionals. Fig. \ref{figure2} shows the
interaction energy curves of water on the BN substrates from PBE, PBE+D3,
PBE0+MBD, optB86b-vdW and SCAN \xc functionals for comparison. We consider two
important aspects when assessing these \xc functionals. First, we compare the
absolute adsorption energy of water on each substrate to the reference methods.
More specifically for the water/h-BN system, we compare to the long-range
corrected adsorption energies in Table \ref{table1}. Second, we consider the
relative trend of the water adsorption energy from the small molecule to the
extended BN surface, keeping in mind that LRDMC and other reference methods
predict that the interaction of water is 10-15 meV less on boronene and h-BN
compared to borazine.

Starting with the most widely used \xc functional, PBE underestimates the
interaction energy in both the molecular systems and on the extended h-BN
surface by as much as $50\%$. This is expected since with this functional, vdW
interactions are not treated.  Note that the hybrid functional PBE0 is not shown
in Fig. \ref{figure2} because it was found to overlap with PBE in the
water/borazine and water/boronene systems, as well as being within 10 meV of PBE
in the water/h-BN system. Despite severely underestimating the adsorption
energies, PBE and PBE0 correctly predict that the water/boronene interaction
energy is less than the water/borazine interaction energy.

Dispersion inclusive \xc functionals such as optB86b-vdW and vdW-DF2 are
generally considered as appropriate methods for predicting the properties of
layered materials and vdW dominated complexes. From Fig. \ref{figure2} it can be
seen that optB86b-vdW and vdW-DF2 provide good agreement with reference
interaction energies for water/borazine near the minimum at 3.32 \AA, predicting
interaction energies of $-126$ meV and $-118$ meV, respectively. However, these
dispersion inclusive functionals estimate a $20-40$ meV stronger interaction for
the water/boronene and water/h-BN systems compared to reference interaction
energies. As such, optB86b-vdW and vdW-DF2 predict the wrong trend for the
adsorption energy from zero to 2D systems, overestimating the adsorption energy
at the minimum by up to $40\%$ on the larger BN substrates. Indeed, similar
behavior has been recently seen for water inside a carbon
nanotube\cite{al-hamdani4}. In Section \ref{tddft} we will elaborate on this
overestimation.

Dispersion interactions can also be accounted for in DFT calculations with
dispersion corrections to \xc functionals, such as in PBE+D3. Although PBE+D3
can be seen to overestimate the interaction energy of water on all of the BN
substrates in Fig. \ref{figure2}, it predicts, in agreement with the reference
methods, that the adsorption energy does not vary significantly from zero to 2D.
Let us also consider the MBD correction which has been evaluated in combination
with PBE0 for different supercells in Table \ref{table2} and the interaction
energy curves are shown in Fig. \ref{figure2}. In this way the periodic dipole
potential is summed over a long distance and all possible collective charge
density fluctuations are converged. Indeed, PBE0+MBD captures the same
higher-order, non-additive correlation interactions as the RPA. However, the MBD
correction with PBE0 leads to 20-30$\%$ overestimation of the interaction energy
on the molecular and surface substrates, compared to explicitly correlated
methods.

The best performance amongst the \xc functionals we have considered is given by
the recently developed SCAN functional. SCAN predicts the water/borazine
interaction energy to be $-126$ meV at a distance of 3.22 \AA\ and it agrees
perfectly with DMC at water-borazine separations above 3.5 \AA. In the absence
of any dispersion correction to SCAN, the interaction energy at larger
water-borazine separation distances reflects the LDA-like construction of this
functional and the inclusion of some long-range correlation energy that was
incorporated using an Ar dimer interaction\cite{scan}. SCAN also predicts a 15
meV weaker interaction for water/boronene compared to water/borazine - in line
with the trends given by LRDMC, DMC, LCCSD(T) and MP2. However, SCAN predicts
interaction energies in close agreement with the reference interaction energies
of water/boronene and water/h-BN, despite the absence of any vdW correction.
Thus, the implementation of any vdW-correction to SCAN has to be done cautiously
in order to avoid considerable errors. Still, it has been shown that many
properties derived by SCAN can be improved by incorporating a long-range
dispersion correction.\cite{scand3} For a detailed analysis of the performance
of a selection of other widely used \xc functionals on the h-BN surface, we
refer the interested reader to Ref. \citenum{al-hamdani2}.

In this section we have seen that various DFT \xc functionals need further
improvement, and the challenge arises in the larger systems for which the
inclusion of vdW interactions leads to an overestimation of the interaction
energy. The likely sources of error in the \xc functionals are discussed in
Section \ref{tddft}.

\subsection{Static and dynamic polarizabilities of the water/BN interactions
from TD-DFT}\label{tddft} The adsorption of water on BN substrates of different
sizes and dimensions is found to have almost the same adsorption energy, as
demonstrated in Fig. \ref{figure_prop}a. This is contrary to the naive
expectation that the adsorption energy increases with the size of the system due
to increased dispersion interaction. The expected behavior of the water
adsorption energy is given instead by vdW models such as optB86b-vdW and
vdW-DF2, also shown in Fig. \ref{figure_prop}a. These methods predict a
monotonic increase in the adsorption energy of water with the increasing size of
the BN substrate. In order to understand this outcome better, the static and
dynamic polarlizabilities of each system has been computed. These
polarizabilities are observable and therefore provide information that is useful
to both computational and experimental studies. The static polarizability
$\alpha_{kl}^{0}$ of a system describes its response to external fields giving
rise to the induction energy.
A change in the induction energy will be mainly determined by the static
polarizability of the BN systems. We report the in-plane and out-of-plane
polarizabilities \alphiso~ and \alphfull, respectively, where the latter is
particularly relevant for the present adsorption geometries. The dynamic
polarizability $\alpha_{kl}(i\omega)$, on the other hand, determines the
dispersion interaction in a system, its leading order can be described by
molecular \csix\ dispersion coefficients.\cite{stone1997,casimir1948} We present
these intermolecular \csix\ coefficients for the interaction between water and
the different BN systems. A common approximation is the use of isotropic dynamic
polarizabilities
($\alpha_{\text{iso}}=\frac{1}{3}\text{Tr}\left[\alpha_{\text{kl}}\right]$).
This results in an isotropic dispersion coefficient \ciso, which we contrast
with \cfull arising from the full polarizability tensor. Note that the above
description of induction and dispersion interaction is only valid in the
long-range (\textit{i.e.}  well separated) limit.

The molecular polarizabilities $\alpha$ have been computed from TD-DFT with well
established numerical settings (PBE38 functional in def2-QZVPD-aug basis,
solving a non-standard eigenvalue problem in frequency
domain).\cite{dftd3,casida1995} A comparison to the MBD model polarizabilities
is shown in Fig. \ref{figure_prop}. Details of the TD-DFT calculations and the
polarizability evaluations can be found in the SI along with the structures of
the water/BN complexes.
\begin{figure}[ht] \centering
\includegraphics[width=0.5\textwidth]{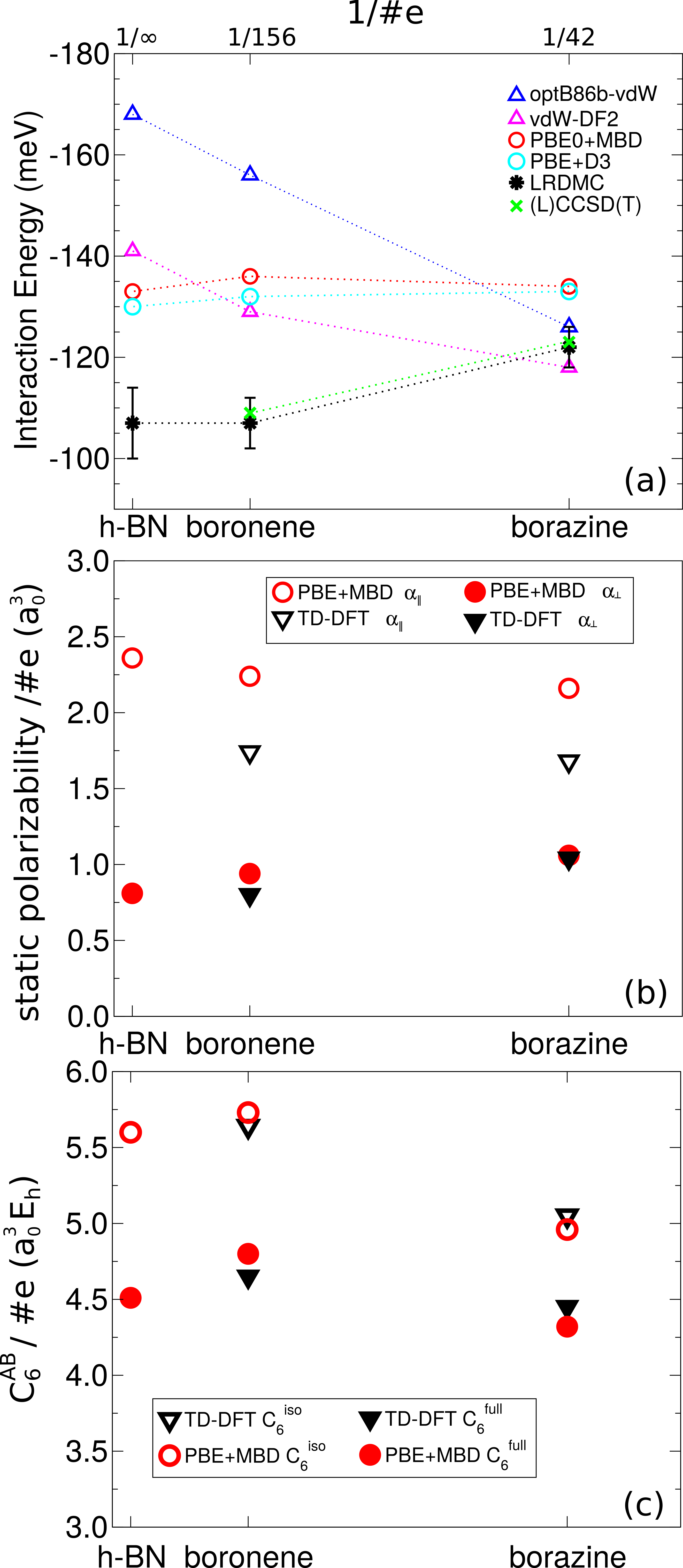}
\caption{Observable properties of water/BN systems. (a) Adsorption energy in meV
of water on BN substrates: h-BN with ($4\times4$) unit cell, boronene, and borazine.
(b) Static polarizability normalised for the total number of electrons in the BN substrate,
from TD-DFT (black triangles) and PBE+MBD (red circles). Empty symbols indicate
the in-plane static polarizability \alphiso. Filled symbols indicate the
out-of-plane static polarizability \alphfull. (c) Molecular \ciso\ and \cfull\
per electron with respect to the BN substrate, from TD-DFT (black triangles) and
PBE+MBD (red circles).~\ciso\ is the isotropic dynamic polarizability and is
indicated by empty symbols.~\cfull\ is calculated from the full dynamic
polarizability tensor accounting for anisotropy and is indicated by filled symbols.}
\label{figure_prop} \end{figure}
\begin{table}[tbp]
    \begin{ruledtabular}
      \centering
      \caption{Reference information for the polarizabilities of
        water/BN systems: In-plane (\alphiso) and out-of-plane (\alphfull)
        static polarizabilities [$a_0^3$], and isotropic (\ciso) and full
        (\cfull) dispersion coefficients [$E_h\,a_0^6$] from TD-DFT and PBE0+MBD.
        All values are given per electron and $a_0$ is Bohr's radius.}
      \label{table3}
\begin{tabular}{lrrrr}
  & \alphiso & \alphfull & \ciso & \cfull \\ \hline
\textit{TD-DFT} & & & & \\
borazine          & 1.68     & 1.04      & 5.05  & 4.45   \\
boronene          & 1.74     & 0.80      & 5.64  & 4.65   \\
\textit{PBE0+MBD} &  &  & & \\
borazine          & 2.16     & 1.06      & 4.96  & 4.32   \\
boronene          & 2.24     & 0.94      & 5.73  & 4.80   \\
h-BN              & 2.36     & 0.81      & 5.60  & 4.51
\end{tabular}
\end{ruledtabular}
\end{table}

Let us start by considering the static polarizabilities. Fig. \ref{figure_prop}b
shows that \alphiso\ is significantly higher than \alphfull\ on all BN
substrates, and also invariant with respect to the substrate. However,
\alphfull\ is reduced from borazine to boronene. As such, anisotropy is clearly
important in this system, not only resulting in a reduced induction energy for
each water/BN system, but also changing the relative trend. The neglect of
anisotropy leads to an overestimation of the static polarizability, indicating
that the induction interaction with water is also overestimated. Note that this
form of anisotropy (or non-additivity) is captured at the canonical HF level of
theory, as well as by \xc functionals in DFT. However, it is not given that the
absolute value of the static polarizability is accurate within either of these
approaches and is unlikely to be captured by classical force fields.

Second, the effective molecular \csix\ coefficients have been computed for
water/borazine and water/boronene from TD-DFT. The TD-DFT results show that
\cfull\ is $14\%$ smaller than \ciso\ for water/borazine, and $21\%$ smaller for
water/boronene. Once again, this suggests that anisotropy plays a key role in
the dispersion interaction, such that the dispersion energy is overestimated in
the isotropic case. Indeed, in Fig. \ref{figure_prop}a the predicted adsorption
energies on boronene and h-BN from optB86b-vdW and vdW-DF2 demonstrate that an
isotropic model of vdW interactions leads to a significantly larger estimation
of the adsorption energy of water.

PBE+MBD as well as PBE+D3 have been found to overestimate the interaction
energies despite reproducing the stability of the water adsorption energy from
zero to 2D. The effective molecular \csix\ coefficients are used to assess the
accuracy of MBD for predicting the dispersion contribution to the interaction
energies in the molecular adsorption systems. In fact, we can see from Table
\ref{table3} that MBD accurately predicts \cfull\ within 3$\%$ of the reference
TD-DFT for water/borazine and water/boronene and we can thus trust the
water/h-BN results. The static polarizability for h-BN is comparable, though
slightly smaller than for boronene. Although the absolute \csix\ coefficient is
also larger on h-BN than on boronene, after normalising for the number of
electrons, \cfull\ appears slightly smaller on the extended BN surface. The
small difference can be attributed to the lack of hydrogen atoms in the h-BN
surface, which contribute to \cfull\ in boronene. As the long-range limit in
PBE-MBD seems to be well captured, the short to medium range interaction is more
likely to be a source of error and in particular, the interface between MBD and
the underlying \xc functional.

To summarize the findings in this section, molecular polarizabilities of
water/borazine, water/boronene, and water/h-BN systems have been computed using
TD-DFT and MBD. The results of the static polarizability have indicated that the
induction energy decreases from the small molecular borazine system to the
larger boronene molecule and to the extended h-BN surface. At the same time it
is countered by the increase in the dispersion interaction, resulting in almost
the same adsorption energy of water on these BN substrates that span zero to
2-dimensions. Furthermore, the anisotropy in these systems is significant and
cannot be captured using isotropic dispersion coefficients or standard vdW
models.

\section{Conclusion}\label{conc} The adsorption energy of water on BN substrates
has been determined from coupled cluster theory, LRDMC, DMC, MP2 and RPA based
methods. The best estimate of the water/h-BN adsorption energy is $-107\pm7$ meV
from LRDMC. Various corrections have been quantified and the most significant
include: single excitations to the RPA, perturbative triple excitations in
CCSD(T), and the contribution from long range correlation energy on the h-BN
surface. Each was found to contribute $\sim$20 meV to the adsorption energy of
water.
We report static polarizabilities and effective \csix~ dispersion coefficients
from TD-DFT and MBD. Interestingly, a significant amount of non-additivity is
found in the dispersion interaction of water with h-BN, despite the substrate
being a wide band gap insulating system. The non-additive interaction in these
systems is due to the high degree of anisotropy. The findings show that the MBD
correction is a promising method for estimating the long-range correlation
contribution especially for highly anisotropic, low-dimensional structures. This
is particularly useful for more expensive many-body wavefunction based periodic
methods, in which some of the inherent finite size effects can be estimated by
the MBD method in future. However, it is clear for the \xc functionals
considered here that there is still a lot of scope for improvement --
particularly in the development of \xc functionals for the accurate prediction
of adsorption energies. The adsorption energies, static polarizabilities and
molecular dispersion coefficients presented in this study also provide an
opportunity for future experimental measurements of these properties to be
compared.

\acknowledgments We are grateful for support from University College
London and Argonne National Laboratory (ANL) through the Thomas Young
Centre-ANL initiative. Some of the research leading to these results
has received funding from the European Research Council under the
European Union's Seventh Framework Programme (FP/2007-2013) / ERC
Grant Agreement number 616121 (HeteroIce project). J.G.B. acknowledges
support by the Humboldt Foundation within the Feodor-Lynen program.
A.M. is supported by the Royal Society through a Wolfson Research
Merit Award. This research used resources of the Argonne Leadership
Computing Facility at Argonne National Laboratory, which is supported
by the Office of Science of the U.S. DOE under contract
DE-AC02-06CH11357. This research also used resources as part of an
INCITE project (awarded to D.A.)  at the Oak Ridge National Laboratory
(Rhea/Eos), which is supported by the Office of Science of the
U.S. Department of Energy (DOE) under Contract No.
DEAC05-00OR22725. In addition, we are grateful for computing resources
provided by the London Centre for Nanotechnology and research
computing at University College London. M.R. acknowledges support from
the University of Oxford and the {\'E}cole Polytechnique
F{\'e}d{\'e}rale de Lausanne, as well as funding from the Otto Hahn
Award of the Max Planck Society.  B.R. and G.K. acknowledge funding by
the Austrian Science Fund (FWF) F41 (SFB ViCoM).  Computations were
performed on the Vienna Scientific Cluster VSC3.

\newpage

\setcounter{section}{0}

\renewcommand{\thesection}{S\arabic{section}}%

\setcounter{table}{0}

\renewcommand{\thetable}{S\arabic{table}}%

\setcounter{figure}{0}

\renewcommand{\thefigure}{S\arabic{figure}}%

\section*{Supporting Information}
In this material we give further details regarding the
calculations of the static and dynamic polarizabilities of water-BN
complexes. In addition, the interaction of water with BN substrates
has been computed for specific geometries which we report here for the
complexes near the minimum and separated far apart.

\subsection{Computing Static and Dynamic Polarizabilities}
To compute the intermolecular C6AB coefficients, we use the second
order perturbation expression of the dispersion energy, e.g. as given
Ref.\citenum{stone} Eq. 4.3.9.
\begin{equation}
E^AB(6,disp)=-1/(2\pi) T_{ab}T_{cd}
\int [ \alpha^A_{ac}(iw) \alpha_{bd}(iw) ]dw =! -C_6^{AB} / R^6
\end{equation}
with the dipole operators $T_{ab}=(3R_aR_b-R^2 \delta_{ab})/R^5$ and
the dynamic polarizabilities $\alpha$. For the considered system
(borazine/boronene in x-y plane, water displaced along z-axis),
$R_x=R_y=0$ and $R_z=R$. The polarizability tensors only have diagonal
non-zero contributions, leading to the dipole-dipole response
\begin{equation}
C_6^{AB} = 1/(2\pi) \int [ \alpha^A_{xx}(iw)\alpha^B_{xx}(iw) +
  \alpha^A_{yy}(iw)\alpha^B_{yy}(iw) +
  4\alpha^A_{zz}(iw)\alpha^B_{zz}(iw) ] dw
\end{equation}
or in the isotropic
approximation,
\begin{equation}
\alpha_{xx}=\alpha_{yy}=\alpha_{zz}=1/3 Tr[\alpha]
C_6^{AB}=3/\pi \int [ \alpha^A(iw)\alpha^B(iw) ] dw
\end{equation}

Dynamic polazizability tensors are computed with the TURBOMOLE program
package\cite{turbomole} employing a modification of the PBE0
functional\cite{pbe0} with a fraction of 3/8 of Fock exchange. The single
particle orbitals are expanded in a large Gaussian basis set of quadruple-zeta
quality def2-QZVP\cite{def2basen} with additional flat functions (2s2p on H,
2s2pd on C, O, B, N). The dynamical polarizability is computed for a set of
frequency points up to $\omega=272$eV. The required frequency integration is
performed numerically by interpolating the frequency points with cubic splines
and extrapolating with an exponentially decaying function above 272eV. This
setting has been established and tested previously.\cite{dftd3}

\subsection{Geometries of water-BN complexes}

Here, one can find the geometries for water/borazine and
water/boronene configurations in xyz file format with cartesian
coordinates. The water/h-BN configuration is also given in VASP file
format, showing the structure as well as the unit cell details.

\begin{verbatim}
15
water-borazine complex at 3.32 AA
H   5.02076127599995   7.64373657450001   0.60758282549997
H   7.10527023150000   6.17375954850001   0.58709578200002
H   7.28364120750005   3.63461177700003   0.41760386850004
H   4.96865547299997   2.56629642749996   0.32842254299997
H   2.67920894549999   3.68622618450004   0.36779496749997
H   2.90474809949995   6.22044015000000   0.50692081349997
H   5.29230349199999   4.92565706249998   4.09910000000000
H   6.08425933200003   5.49983635649998   2.90580000000000
O   6.07642824900005   5.45929539300001   3.88230000000000
B   5.00435273549996   6.44422098149999   0.55828330799997
B   3.72565904700004   4.27443283650000   0.41692162499995
B   6.24820422149998   4.24146091050005   0.45255167849998
N   6.21955164299997   5.67465859050003   0.56921574450005
N   4.98001838850001   3.57987289349996   0.39688556550001
N   3.77999653349999   5.70466272150004   0.49999999950000
\end{verbatim}
\begin{verbatim}
15
water-borazine complex at 7.5 AA
H  5.0207612760   7.6437365745   0.6075828255   
H  7.1052702315   6.1737595485   0.5870957820  
H  7.2836412075   3.6346117770   0.4176038685  
H  4.9686554730   2.5662964275   0.3284225430  
H  2.6792089455   3.6862261845   0.3677949675  
H  2.9047480995   6.2204401500   0.5069208135  
H  5.2923034920   4.9256570625  8.19910000000 	      
H  6.0842593320   5.4998363565  7.00580000000 	      
O  6.0764282490   5.4592953930  7.98230000000 	      
B  5.0043527355   6.4442209815   0.5582833080  
B  3.7256590470   4.2744328365   0.4169216250  
B  6.2482042215   4.2414609105   0.4525516785  
N  6.2195516430   5.6746585905   0.5692157445  
N  4.9800183885   3.5798728935   0.3968855655  
N  3.7799965335   5.7046627215   0.4999999995
\end{verbatim}

\begin{verbatim}
39
water/boronene complex at 3.40 AA
H   14.91534369999997   11.41999646599999   4.99999357280001
H   13.45245660200007   13.46476483399996   5.00098993280005
H   13.57405612999997   6.77571078000000   4.99997236959996
H   14.96153026599998   8.87207758200006   5.00013619039999
H   8.87735747999997   5.20870275199997   4.99816014240002
H   11.37952545399997   5.45439400000006   4.99990587840000
H   5.47960577200001   11.25596668000000   5.00407952319995
H   6.51897141399999   13.54427477200005   4.99912300480003
H   6.64760796200000   6.44317854999997   4.99906735039997
H   5.52603312599999   8.69262581000001   5.00446285440000
H   11.21154280799999   14.70503203800007   5.00087987200004
H   8.70191243799994   14.85927119999999   4.99804943679999
H   8.97609295600006   9.99796941200003   7.47900000000004
H   10.24453276000003   9.74192531200003   8.31500000000005
O   9.29857606200002   9.95683262200004   8.39999999999998
B   7.24498310399994   7.48508810999994   5.00000000000000
B   9.40077044600002   8.78218644599997   5.00000000000000
B   7.19930123400005   10.00518914599994   5.00000000000000
B   9.35591138199996   11.30765023799995   5.00000000000000
B   11.55712669400003   10.08520101200006   5.00000000000000
B   11.50858032200006   12.59770345800007   5.00000000000000
B   9.30608056599993   13.82112965199994   5.00000000000000
B   7.15414683199995   12.52510048199994   5.00000000000000
B   9.44278340400004   6.26835964400001   5.00000000000000
B   11.59976091000004   7.57104182800006   5.00000000000000
B   13.76018130800006   8.86851596000000   5.00000000000000
B   13.71470368000004   11.38052640799998   5.00000000000000
N   8.69419132999994   7.50934365999996   5.00000000000000
N   8.65557782799996   10.03167293000004   5.00000000000000
N   10.80548743600005   11.32588335400001   5.00000000000000
N   6.54226394800006   8.72565721399994   5.00000000000000
N   10.85067710999994   8.81655632999994   5.00000000000000
N   8.60288796399999   12.55350658599994   5.00000000000000
N   10.73135179800005   13.80887621599996   5.00000000000000
N   10.86695565799999   6.33242285199998   5.00000000000000
N   6.49634705599993   11.25977556999999   5.00000000000000
N   13.03791568799994   7.63958597200002   5.00000000000000
N   13.01480868599995   10.11107383600006   5.00000000000000
N   12.94844807800004   12.58176827599996   5.00000000000000
\end{verbatim}
\begin{verbatim}
39
water/boronene complex at 8.00 AA
H   14.91534369999997   11.41999646599999   4.99999357280001
H   13.45245660200007   13.46476483399996   5.00098993280005
H   13.57405612999997   6.77571078000000   4.99997236959996
H   14.96153026599998   8.87207758200006   5.00013619039999
H   8.87735747999997   5.20870275199997   4.99816014240002
H   11.37952545399997   5.45439400000006   4.99990587840000
H   5.47960577200001   11.25596668000000   5.00407952319995
H   6.51897141399999   13.54427477200005   4.99912300480003
H   6.64760796200000   6.44317854999997   4.99906735039997
H   5.52603312599999   8.69262581000001   5.00446285440000
H   11.21154280799999   14.70503203800007   5.00087987200004
H   8.70191243799994   14.85927119999999   4.99804943679999
H   8.97609295600006   9.99796941200003   12.07899999999995
H   10.24453276000003   9.74192531200003   12.91499999999996
O   9.29857606200002   9.95683262200004   13.00000000000000
B   7.24498310399994   7.48508810999994   5.00000000000000
B   9.40077044600002   8.78218644599997   5.00000000000000
B   7.19930123400005   10.00518914599994   5.00000000000000
B   9.35591138199996   11.30765023799995   5.00000000000000
B   11.55712669400003   10.08520101200006   5.00000000000000
B   11.50858032200006   12.59770345800007   5.00000000000000
B   9.30608056599993   13.82112965199994   5.00000000000000
B   7.15414683199995   12.52510048199994   5.00000000000000
B   9.44278340400004   6.26835964400001   5.00000000000000
B   11.59976091000004   7.57104182800006   5.00000000000000
B   13.76018130800006   8.86851596000000   5.00000000000000
B   13.71470368000004   11.38052640799998   5.00000000000000
N   8.69419132999994   7.50934365999996   5.00000000000000
N   8.65557782799996   10.03167293000004   5.00000000000000
N   10.80548743600005   11.32588335400001   5.00000000000000
N   6.54226394800006   8.72565721399994   5.00000000000000
N   10.85067710999994   8.81655632999994   5.00000000000000
N   8.60288796399999   12.55350658599994   5.00000000000000
N   10.73135179800005   13.80887621599996   5.00000000000000
N   10.86695565799999   6.33242285199998   5.00000000000000
N   6.49634705599993   11.25977556999999   5.00000000000000
N   13.03791568799994   7.63958597200002   5.00000000000000
N   13.01480868599995   10.11107383600006   5.00000000000000
N   12.94844807800004   12.58176827599996   5.00000000000000
\end{verbatim}

\begin{verbatim}
water/h-BN complex at 3.20 AA                         
1.0
 10.0472469085   0.0000379684  -0.0003311695 
  5.0238219661   8.7014468612   0.0002996894 
  0.0000000000   0.0000000000  16.0000000000 
H O B N
 2 1 16 16
Cartesian
  5.8848699130   3.4640408839  5.71700000000
  6.3939727004   3.6273612958  4.26700000000
  6.6644305907   3.7324806797  5.20000000000
  1.2944720000   0.6967070000   1.9996680000
  2.5504565567   2.8716975549   1.9995382176
  3.8064175128   5.0469969904   2.0003546472
  5.0624350751   7.2223780438   2.0014071730
  3.8062926218   0.6962529135   2.0001958196
  5.0622540727   2.8715620758   2.0014348620
  6.3182004626   5.0469377116   2.0028089602
  7.5740702489   7.2224434846   2.0017055725
  6.3181019094   0.6961900093   2.0015271752
  7.5740099987   2.8715106653   2.0028458212
  8.8298994553   5.0470284660   2.0023236716
 10.0859240046   7.2223339056   2.0008677755
  8.8298671810   0.6961228629   2.0013495380
 10.0857545005   2.8716831002   2.0005057810
 11.3417709433   5.0469911404   1.9996170630
 12.5977645849   7.2222727337   2.0006252347
  2.5504528649   1.4215297352   2.0001858400
  3.8063819878   3.5967701297   2.0012485940
  5.0623413269   5.7721038675   2.0026259662
  6.3182617717   7.9477122224   2.0016795990
  5.0622048194   1.4213476926   2.0020270554
  6.3181514198   3.5966695226   2.0043775345
  7.5740485050   5.7721773001   2.0043577256
  8.8300307416   7.9474750657   2.0017316786
  7.5740075331   1.4212914889   2.0034872407
  8.8298678785   3.5967699709   2.0035198381
 10.0858659396   5.7720851029   2.0016183961
 11.3418803526   7.9474067513   2.0017782839
 10.0857404189   1.4214876504   2.0009212127
 11.3417145387   3.5967752907   2.0001414713
 12.5977078237   5.7720493736   2.0001252461
 13.8537300000   7.9472590000   2.0023270000
\end{verbatim}
\begin{verbatim}
water-hBN                                
1.0
 10.0472469085   0.0000379684  -0.0003311695 
  5.0238219661   8.7014468612   0.0002996894 
  0.0000000000   0.0000000000  16.0000000000 
H O B N
 2 1 16 16
Cartesian
  5.8848699130   3.4640408839  10.5170000000 
  6.3939727004   3.6273612958  9.06700000000 
  6.6644305907   3.7324806797  10.0000000000 
  1.2944720000   0.6967070000   1.9996680000 
  2.5504565567   2.8716975549   1.9995382176 
  3.8064175128   5.0469969904   2.0003546472 
  5.0624350751   7.2223780438   2.0014071730 
  3.8062926218   0.6962529135   2.0001958196 
  5.0622540727   2.8715620758   2.0014348620 
  6.3182004626   5.0469377116   2.0028089602 
  7.5740702489   7.2224434846   2.0017055725 
  6.3181019094   0.6961900093   2.0015271752 
  7.5740099987   2.8715106653   2.0028458212 
  8.8298994553   5.0470284660   2.0023236716 
 10.0859240046   7.2223339056   2.0008677755 
  8.8298671810   0.6961228629   2.0013495380 
 10.0857545005   2.8716831002   2.0005057810 
 11.3417709433   5.0469911404   1.9996170630 
 12.5977645849   7.2222727337   2.0006252347 
  2.5504528649   1.4215297352   2.0001858400 
  3.8063819878   3.5967701297   2.0012485940 
  5.0623413269   5.7721038675   2.0026259662 
  6.3182617717   7.9477122224   2.0016795990 
  5.0622048194   1.4213476926   2.0020270554 
  6.3181514198   3.5966695226   2.0043775345 
  7.5740485050   5.7721773001   2.0043577256 
  8.8300307416   7.9474750657   2.0017316786 
  7.5740075331   1.4212914889   2.0034872407 
  8.8298678785   3.5967699709   2.0035198381 
 10.0858659396   5.7720851029   2.0016183961 
 11.3418803526   7.9474067513   2.0017782839 
 10.0857404189   1.4214876504   2.0009212127 
 11.3417145387   3.5967752907   2.0001414713 
 12.5977078237   5.7720493736   2.0001252461 
 13.8537300000   7.9472590000   2.0023270000 

\end{verbatim}

\end{document}